\documentclass{PoS}

\title{B-physics with $N_f=2$ Wilson fermions}

\ShortTitle{B-physics with $N_f=2$ Wilson fermions}

\author{\speaker{F.~Bernardoni}$\,^{a}$, B.~Blossier$\,^{b}$,
J.~Bulava$\,^{c}$, M.~Della\ Morte$\,^{h}$, P.~Fritzsch$\,^{e}$,
N.~Garron$\,^{f}$, A.~G\'erardin$\,^{b}$, J.~Heitger$\,^{g}$,
G.~von~Hippel$\,^{d}$, H.~Simma$\,^{a}$, R.~Sommer$\,^{a}$\\
$^a$~NIC, DESY, Platanenallee~6, 15738~Zeuthen, Germany\\
$^b$~Laboratoire~de~Physique~Th\'eorique, CNRS/Universit\'e~Paris~XI,
F-91405~Orsay~Cedex, France\\
$^c$~CERN, Physics~Department, TH~Division, CH-1211~Geneva~23, Switzerland\\
$^d$~Institut~f\"ur~Kernphysik, University~of~Mainz, Becher-Weg~45, 55099~Mainz, Germany\\
$^e$~Institut~f\"ur~Physik, Humboldt-Universit\"at~zu~Berlin, Newtonstr.~15, 12489~Berlin, Germany\\
$^f$~School~of~Mathematics, Trinity~College, Dublin~2, Ireland\\
$^g$~Universit\"at~M\"unster, Institut~f\"ur~Theoretische~Physik, Wilhelm-Klemm-Str.~9, 48149~M\"unster,
Germany\\
$^h$~IFIC and CSIC, Calle Catedr\'atico Jos\'e Beltran, 2, 46980 Paterna, Valencia, Spain
\\
\email{fabio.bernardoni@desy.de}}

\abstract{
\flushright{SFB/CPP-13-61\\DESY 13-154\\IFIC/13-58\\MS-TP-13-22\\ HU-EP-13/41\\TCD-MATH-13-10\\}
\vspace{0.4cm}
 We report the final results of the ALPHA collaboration
 for some B-physics observables: $f_B$, $f_{B_s}$ and $m_b$. We employ CLS 
 configurations with 2 flavors of $O(a)$ improved Wilson fermions 
 in the sea and pion masses ranging down to 190 MeV. The b-quark 
 is treated in HQET to order $1/m_b$. The renormalization, the matching 
 and the improvement were performed non-perturbatively, and three lattice 
 spacings reaching $a=0.048$ fm are used in the continuum extrapolation.}

\FullConference{31st International Symposium on Lattice Field Theory LATTICE 2013\\
                 July 29 - August 3, 2013\\
                 Mainz, Germany}

\usepackage[T1]{fontenc}
\usepackage[utf8]{inputenc} 
\usepackage{tikz}
\usepackage{pgf}
\usepackage{pgfplots}
\usepackage{pdfpages}
\usepackage{graphics,graphicx}
\usepackage{wrapfig,booktabs}
\usepackage{macros_static}
\usepackage{amsmath}
\usepackage{amsfonts}
\usepackage{amssymb,bm}
\usepackage{dcolumn,booktabs}
\usetikzlibrary{shapes}
\usetikzlibrary{shadows}
\usetikzlibrary{patterns}
\usetikzlibrary{decorations}
\usetikzlibrary{arrows}
\usetikzlibrary{backgrounds}
\usetikzlibrary{calc}
\usetikzlibrary{fadings}
\usepackage{multirow}
\usepackage{cite}

\begin{document}

\section{Introduction}
Many tests of the Standard Model in the B-physics sector need inputs from the lattice.

First, in experiments such as LHCb and B-factories new physics signals may appear 
through deviations from the Standard Model predictions of B decays, typically 
occurring through the weak interaction. The theoretical computations have some 
hadronic matrix elements as inputs, which due to their intrinsically 
non-perturbative nature can only be computed from first principles on the lattice. 

Second, the b-quark mass (usually in the $\overline{MS}$ scheme) is a 
necessary input in many perturbative computations of SM processes and 
in the running of SM parameters. The most precise determination cited 
by the PDG is obtained by matching the perturbative computation of the 
moments of the cross section $\sigma(e^+e^- \to b\overline{b})$ with experiment. 
In comparison, the lattice computation of the RGI b-quark mass $M_b$ 
by means of step scaling methods, and using the experimental value of the B-meson mass as input, relies on a 
matching with the 2-loop anomalous dimension and 3-loop $\beta$-function in the SF scheme at scales of order 
$\sim100$ GeV. The comparison of these two results obtained with 
completely different systematics provides an interesting test of the Standard Model.

Third, it is an important test of QCD to reproduce the B-mesons mass 
spectrum, which has been measured quite precisely. Due to 
its non-perturbative nature, it can only be computed, in a model independent way,  on the lattice.

The aim of this long-standing project is to provide a computation 
of the above quantities in $N_f=2$ QCD keeping all systematic uncertainties under control.\\
In these proceedings I will review the method employed by the 
ALPHA collaboration to achieve this task and present the results obtained, which I roughly divide into predictions:
\begin{equation}
\label{post} 
f_B,\quad f_{B_s},\quad f_{B_s}/f_B
\end{equation}
and postdictions:
\begin{equation}
\label{pre}
m_b^{\overline{MS}}(m_b) ,\quad m_{B_s}-m_B,\quad m_{B^*}-m_B,\quad m_{B_s^*}-m_{B_s}.
\end{equation}

\section{Setup}

The results presented in this work are obtained from measurements 
on the CLS ensembles detailed in Table \ref{CLSens}, which have 
2 degenerate $O(a)$ improved Wilson quarks in the sea. The improvement
 and renormalization were performed non-perturbatively  
\cite{impr:csw_nf2,DellaMorte:2005se,Fritzsch:2012wq}, and the continuum extrapolation is based 
on three lattice spacings: $ 0.048, 0.065, 0.075\,\,  \mbox{fm}\,,$ 
where the scale was set using $f_{\rm K}$ \cite{Fritzsch:2012wq}.
As it can be seen in Table \ref{CLSens}, the pion masses to be used in 
the chiral extrapolation lie in the range 
$190\, \mbox{MeV}\lesssim \mpi \lesssim 450\,\mbox{MeV}\,,$ 
while volume effects are expected to be sufficiently suppressed in 
$m_\pi L$ ($\sim e^{-m_\pi L}$), given that the condition $m_\pi L \ge 4$ is always satisfied.

The finest lattice corresponds to a cutoff $a^{-1}\sim 4$ GeV and 
therefore cannot be used to simulate the $b$-quark relativistically. 
Our approach is to employ HQET, an effective theory of QCD based on 
an expansion in powers of $1/m_b$. We work at order $1/m_b$ and to 
this order the HQET Lagrangian is:
\begin{align}
 \lag{HQET}(x)
 &= {\cal L}_{h}^{stat}- {\omegakin}  {\cal O}_{kin}(x) 
 - {\omegaspin} {\cal O}_{spin}(x) \\
 &=  \heavyb(x) \,D_0\, \heavy(x) + m_{\rm bare}\heavyb(x)\heavy(x)
 -{\omegakin} \heavyb(x){\bf D}^2\heavy(x) 
 -{\omegaspin} \heavyb(x){\boldsymbol\sigma}\!\cdot\!{\bf B}\heavy(x)\nonumber
\end{align}
\begin{wraptable}{r}{7.5cm}
\begin{tabular}{@{\extracolsep{0.2cm}}ccccc}
\toprule
id  & $L/a$ & $a$ [fm]  &  $m_{\rm PS}$[MeV] & $m_{\rm PS} L$  \\
\midrule     
A4  & $32$  &  0.0748   &$380$  & $4.7$ \\
A5  &       &           &$330$  & $4.0$ \\
B6  & $48$  &           &$270$  & $5.2$ \\ 
\midrule  
E5  & $32$  &  0.0651   &$ 440$ &$4.7$ \\
F6  & $48$  &           &$ 310$ & $5.0$ \\
F7  &       &           &$ 270$ & $4.3$ \\
G8  & $64$  &           &$ 190$ & $4.1$ \\
\midrule
N5  &$48$   &  0.0480   &$440$  & $5.2$ \\
N6  &       &           &$340$  & $4.0$ \\
O7  &$64$   &           &$270$  & $4.2$ \\
\bottomrule
\end{tabular}
\caption{\label{CLSens} CLS ensembles used in this work.}
\end{wraptable}
where $m_{\rm bare}$, $\omegakin$ 
and $\omegaspin$ are parameters to be evaluated 
through a matching with QCD. In particular $m_{\rm bare}$ is needed to absorb the linear divergence 
arising from the $\heavyb(x) \,D_0\, \heavy(x)$ term. The appearance of 
power divergences, which is
    due to operator mixing under renormalization, is a general feature 
of HQET. As a consequence, the renormalization has to be carried out non-perturbatively. 

Notice that the theory remains renormalizable 
at every order in the $1/m_b$ expansion. This is so because the
exponential of the action is to be expanded in powers of $1/m_b$, so that 
the operators suppressed by powers of $m_b$ only appear in correlators as insertions:
\begin{align}
\langle {\cal O} \rangle &= \langle {\cal O} \rangle_{stat} 
+ \omegakin \sum_x \langle {\cal O} {\cal O}_{kin}(x)\rangle_{stat} 
 + \omegaspin  \sum_x \langle {\cal O} {\cal O}_{spin}(x)\rangle_{stat} + O(1/m_b^2)
\\
\langle {\cal O} \rangle_{stat} & = \frac{1}{\cal Z} \int_{fields} {\cal O} \exp 
\left(-a^4\sum_x[{\cal L}_{light}+{\cal L}_{h}^{stat}]  \right) \,.
\end{align}
The advantage of our approach, as compared to other treatments of the $b$-quark, 
such as NRQCD, which are not renormalizable, is that we 
 are able to take the continuum limit, once the power divergencies have
    been subtracted non-perturbatively. In practice 
 this means that we can compute the discretization error 
 on our final results just relying on our data.
 
The method of the ALPHA collaboration for the 
non-perturbative renormalization, matching and 
improvement of HQET has been described in \cite{Blossier:2012qu}. Here we just recall the basic ingredients.
 
The matching with QCD is performed in a small fixed volume 
($L_1\sim 0.4$ fm), such that lattice spacings $a^{-1} \gg m_b$ 
can be simulated while keeping $z_b=M_b L_1 \gg 1$. In order to 
determine the observables in eqs.~\eqref{post} and \eqref{pre} 
without $O(a)$ effects, we need to fix the parameters:
\begin{equation}
\mathbf{\omega}(z) =m_{\rm bare}(z),\,\zahqet(z),\,\cah{1}(z),\, \omegakin(z), \,\omegaspin(z) 
\end{equation}
for a range of $z$ values around $z_b$, where $Z_A^{\rm HQET}$ and $c_A^{(1)}$ are the HQET parameters
    of the renormalized time component of the axial current in NLO HQET. We impose the matching conditions
\begin{equation}
\Phi^{\rm HQET} (z, a) = \Phi^{\rm QCD} (z, 0) = \lim_{a\to 0} \Phi^{\rm QCD} (z, a) \,,
\end{equation}
for a suitable set of observables $\Phi_i$. 
Since we match with $O(a)$ improved QCD in the continuum limit, 
we are able to renormalize and improve HQET in one step.  
Subsequently the $\mathbf{\omega}(z)$ corresponding to the 
lattice spacings to be used in large-volume simulations can be 
determined through a step-scaling procedure. To better control 
discretization effects, 
this process is repeated for two different static actions, denoted by HYP1 and HYP2 \cite{DellaMorte:2003mn}.

These parameters are combined with large-volume 
HQET matrix elements to give physical predictions. 
One important source of systematic errors is then 
the contamination from excited states. If $E_2$ and $E_1$ are the
energies of the first excited state and the ground 
state respectively, this contamination is suppressed by $e^{-(E_2-E_1)t}$, 
where $t$ is the sink-source separation. 
To achieve a better suppression, we solve 
the Generalized Eigenvalue Problem (GEVP)   
for a $3\times 3$ correlator matrix, where 
each entry of the matrix corresponds to a 
different Gaussian smearing level of the B-meson interpolating field. 
In this way we achieve a suppression $e^{-(E_4-E_1)t}$ \cite{Blossier:2009kd}. To ensure 
that the resulting systematic error is negligible we only take 
plateau averages where  $\sigma_{\rm sys}<\frac{1}{3}\sigma_{\rm stat}$ is satisfied,
 with $\sigma_{\rm sys}$ estimated from the GEVP results.

The statistical analysis for all observables presented in the 
following was performed according to the methods described in 
\cite{Schaefer:2010hu}. This allows us to take all correlations 
and autocorrelations into account. In particular the 
errors coming from the fit performed for the continuum and chiral extrapolations 
are included in our statistical error.

\section{Computation of the b-quark mass}

At this point the $\mathbf{\omega}(z)$ are 
known for some values of $z$, namely $z=11,\,13,\,15$, but we need them at 
the physical value for the $b$-quark mass, $z_b$, 
to make physical predictions. To determine $z_b$ we use a physical observable, 
which we chose to be the B-meson mass, $m_B$.

\begin{figure}[h]
\begin{tabular}{cc}
\includegraphics[width=0.5\textwidth]{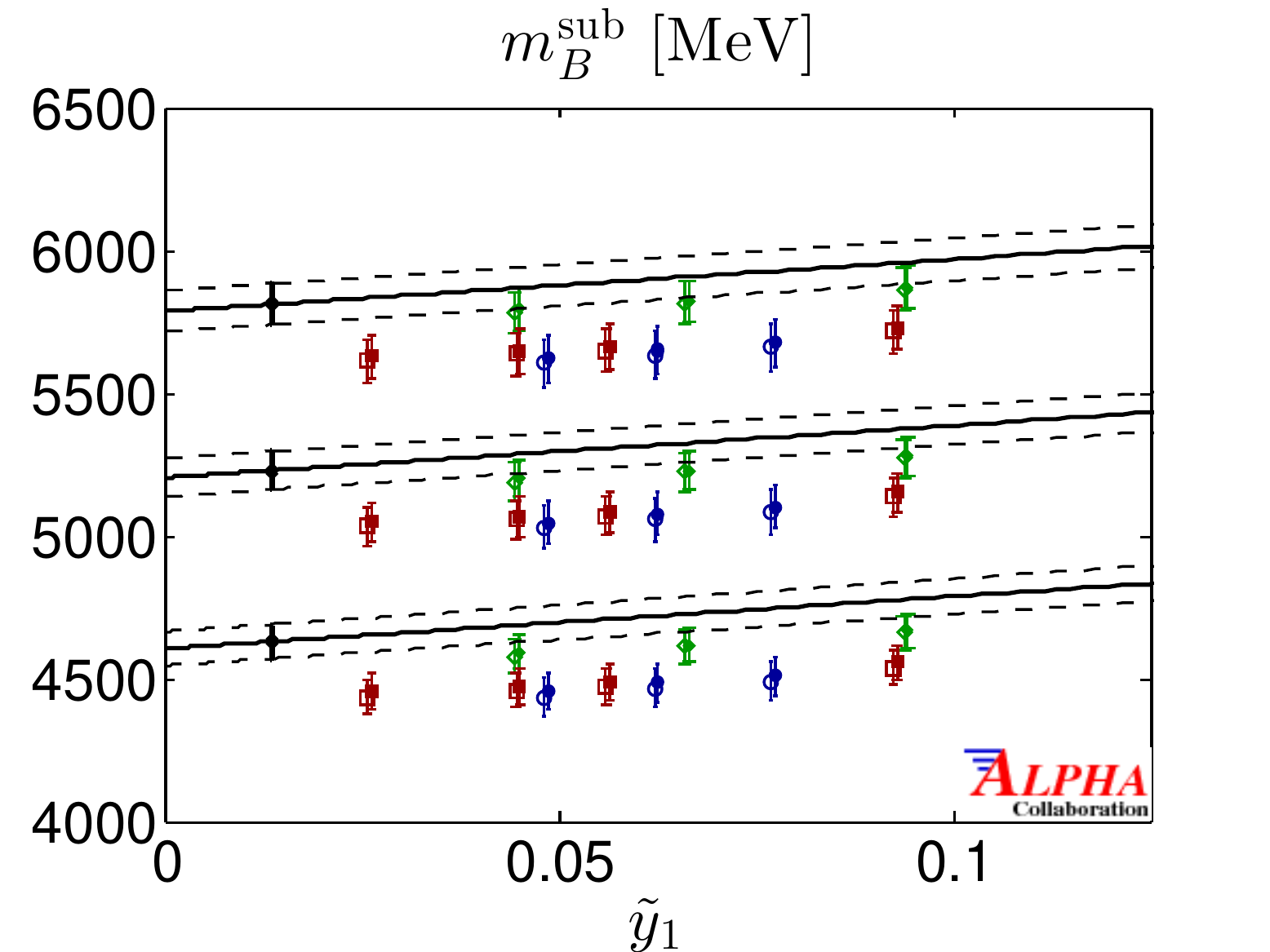}&\includegraphics[width=0.49\textwidth]{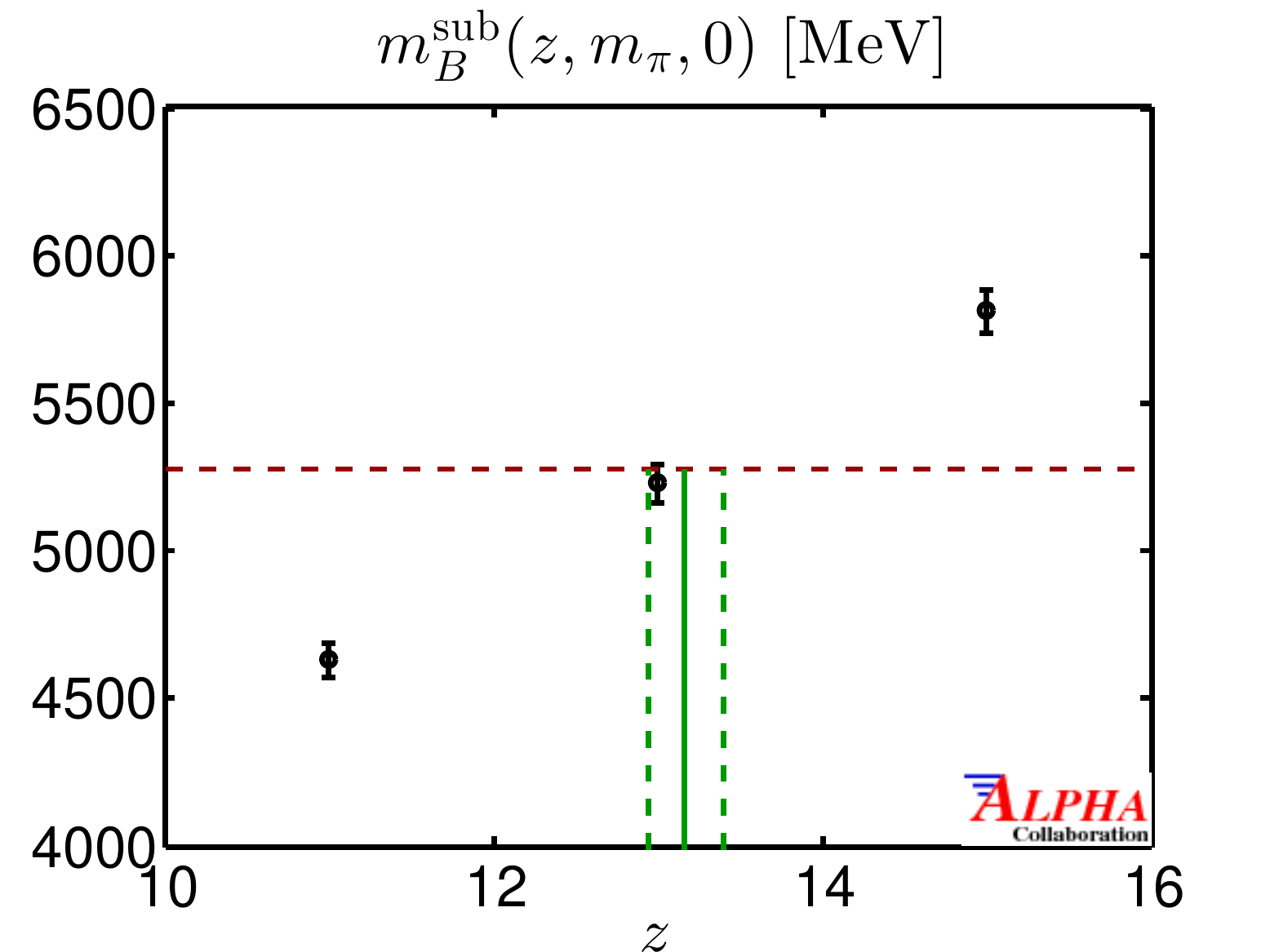}
\end{tabular}
\caption{ \label{mBextrap} Left: Chiral and continuum extrapolation of $m_B^{\rm sub}$. 
Right: $z_b$ determination according to eq.~\protect\eqref{zbfix}. Green, red, blue points correspond to lattice 
spacings $a= 0.048$ fm, $a= 0.065$ fm, and 
$a= 0.078$ fm respectively. Open symbols and 
dashed lines correspond to the HYP1 static action 
while filled symbols and continuum lines correspond to 
HYP2. The band represents the $a=0$ result.}
\end{figure}

We first perform a chiral and continuum extrapolation 
according to the HMChPT formula at NLO 
\cite{Bernardoni:2009sx}\footnote{Here we use the convention $f_\pi\sim 93$ MeV.}:
\begin{align}
m_B^{\rm sub} \equiv m_B\left(z,m_{\rm PS},a,{\rm HYPn}\right) 
+ \frac{3\widehat{g}^2}{32\pi } \left( \frac{m^3_{\rm PS}}{f_{\rm PS}^2}- \frac{m^3_\pi}{f_\pi^2} \right)
= B(z)+ C \left[\tilde{y}_1^{\rm PS}-\tilde{y}_1^\pi \right]
  + D_{\rm HYPn} a^2 \,,
\end{align}
where $\tilde{y}_1^{\rm PS} \equiv \left(\frac{m_{\rm PS}}{4\pi f_{\rm PS}}\right)^2$, 
to get the B-meson mass at the physical pion mass $m_{\rm PS}=m_\pi$ for every $z$. 
For the $B^*B\pi$ coupling in the chiral limit $\widehat{g}$, 
we use the value determined in \cite{Bulava:2010ej}, $\widehat{g}=0.51(2)$. Then we impose:
\begin{equation}
\label{zbfix}
 m_B(z,m_{\pi},a=0)|_{z=z_{\rm b}} \equiv \mB^{\rm exp} =5279.5\MeV
\end{equation}
to find the physical $b$-quark mass. This procedure is illustrated in Fig.~\ref{mBextrap} and we obtain:
\begin{equation}
z_{\rm b} = 13.17(23)(13)_{z} \,.
\end{equation}
The second error arises in the procedure
of fixing $z$. It amounts to $1\%$ and, as explained in \cite{Blossier:2012qu}, it
has to be propagated into QCD observables only after their extrapolation to the
continuum limit.

Using $L_1$ and the $\Lambda$ parameter from \cite{Fritzsch:2012wq} and 
the 4-loop perturbative formulas from 
\cite{vanRitbergen:1997va,Vermaseren:1997fq,Chetyrkin:1997dh,Czakon:2004bu}, 
this value can be converted to the $b$-quark mass in the $\overline{MS}$ scheme. We get:
\begin{equation} 
m_{\rm b}^{\overline{MS}}(m_{\rm b}) = 4.23(11)(3)_{z} \, \mbox{GeV}\,.
\end{equation}

\section{Computation of decay constants and B-meson mass splittings}

We can now compute the $\mathbf{\omega}(z_b)$ 
by quadratically interpolating the  $\mathbf{\omega}(z)$. Combining these parameters with the 
large-volume HQET matrix elements from our ensembles we determine the observables 
of interest up to $O(1/m_b^2)$ effects. 
\begin{figure}
\begin{tabular}{cc}
\hspace{-0.4cm}\includegraphics[width= 0.5\textwidth]{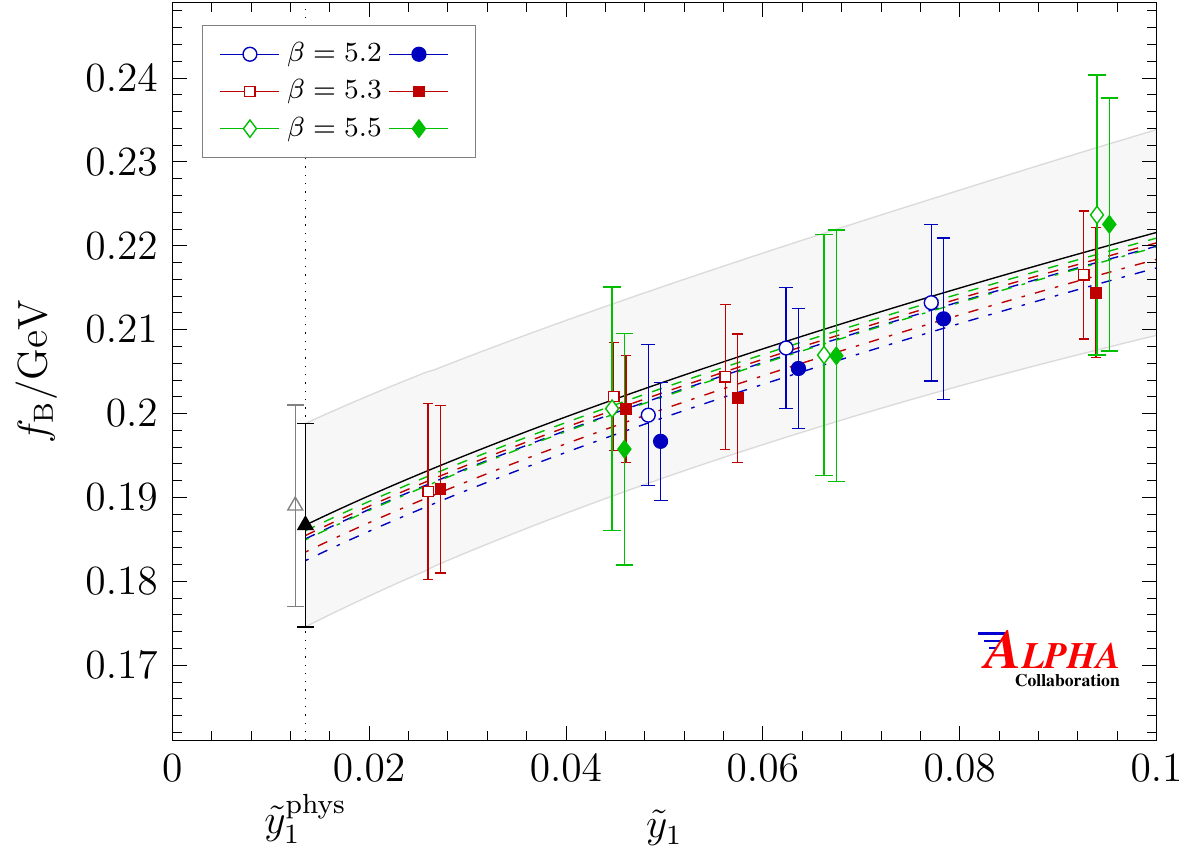}&\includegraphics[width= 0.49\textwidth]{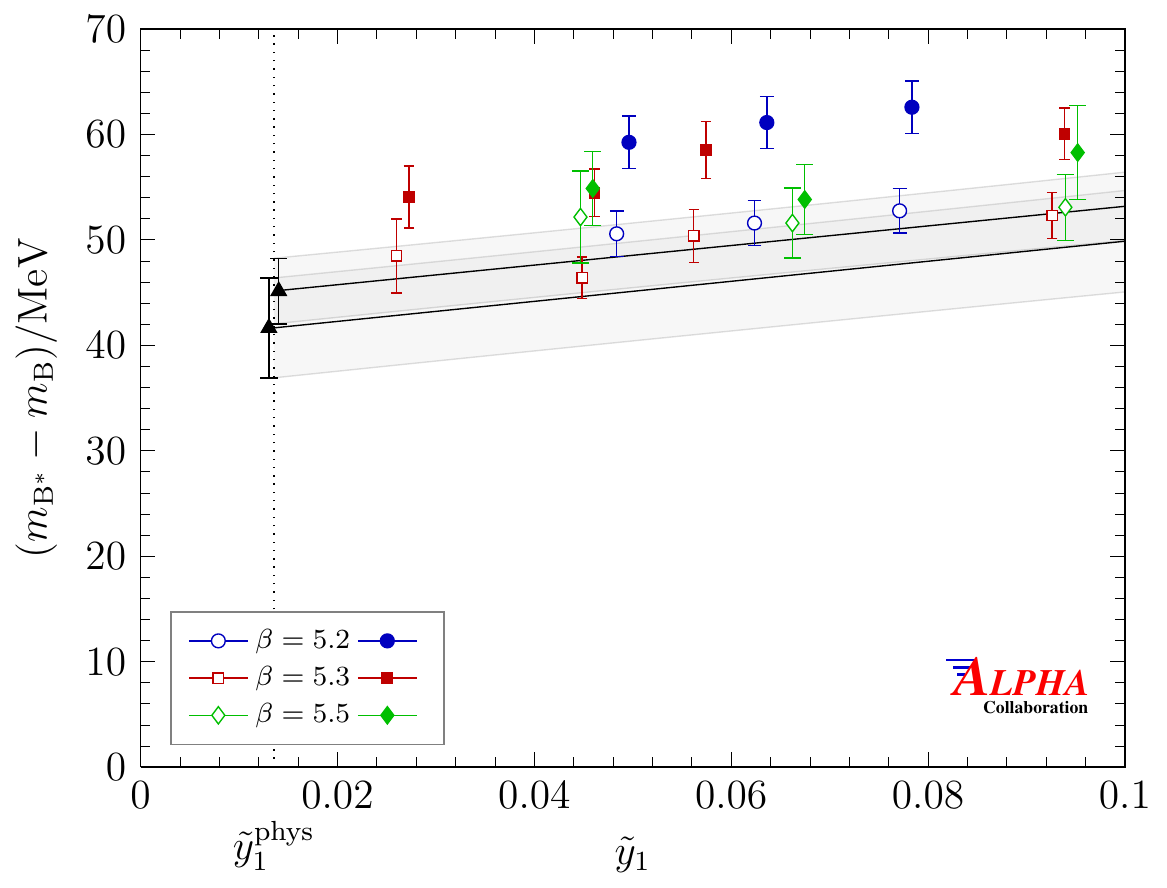}
\end{tabular}
\caption{\label{fBextrap} Combined chiral and continuum extrapolation for $f_B$ (left) and $m_{B^*}-m_B$ (right). 
The same conventions used in Fig.~\protect\ref{mBextrap} apply. 
 On the left, the extrapolation using HMChPT at NLO 
 (filled triangle) is compared with a linear one (open triangle) 
 to extract the systematic error from truncating HMChPT at NLO. 
 On the right, the extrapolations in $a$ and $a^2$ are compared 
 to extract the systematic error due to the continuum extrapolation.}
\end{figure}
Physical results are obtained after a combined chiral and 
continuum extrapolation. Concerning the chiral extrapolations, 
if available we take the NLO HMChPT formulas from 
\cite{Bernardoni:2009sx,Goity:1992tp,Sharpe:1995qp}. Where it is relevant we compare the result from these NLO  extrapolations with
linear ones in $m_{\rm PS}^2$ and quote the difference as a systematic error from the truncation of the chiral
expansion at NLO (subscript $ChPT$). 

Since we neglect effects of order $O(a/m_b)$, 
quantities that are $O(1)$ in HQET are $O(a)$ improved and we expect therefore to reach the 
continuum limit by extrapolating in $a^2$. On the other hand, 
for quantities that are zero in the static limit, such as $m_{B^*}-m_B$, 
effects of $O(a/m_b)$ might still be relevant. We take this into 
account by performing one extrapolation linear in $a$ and one in $a^2$ 
for these quantities and quote the difference as a systematic 
error from the continuum limit (subscript $a$). 
When no systematic error is quoted, it is because 
they are found to be negligible compared to present accuracy.

\begin{table}
\begin{center}
    \begin{tabular}{l|l|l|l}
    Observable & ALPHA & PDG & Method \\ \hline
    $m_B$[MeV]	& input & 5279.5 & $e^+e^-$ scat. \\ \hline
    $m_b^{\overline{MS}}(m_b)$ [GeV] & $4.23(11)(3)_z$ & 4.18(3) & smeared $\sigma(e^+e^- \to b \overline{b})$+
    PT \\ \hline
    $m_{B_s}-m_{B}$  [MeV]& $83.9(6.3)(6.9)_{ChPT}$ & 87.35(0.23) & $pp,\,p\overline{p}$ scat. \\ \hline
    $m_{B^*}-m_{B}$  [MeV]& $41.7(4.7)(3.4)_a$ & 45.3(0.8) & $e^+e^-$ scat. \\ \hline
    $m_{B^*_s}-m_{B_s}$  [MeV]&  $37.9(3.7) (5.9)_a$ & 48.7(2.3) & $e^+e^-$ scat. \\ \hline
    \end{tabular}
    \caption{\label{tabpost} Comparison of ALPHA results for 
    b-quark mass and B-meson mass splittings with Particle Data Group. }
\end{center}
\end{table}

\begin{table}
\begin{center}
    \begin{tabular}{l|l|l|l}
    Obs. & ALPHA &  Lat. Av. & Experiment \\ \hline
    $f_B$ [MeV] & $187(12)(2)_{ChPT}$	& 197(10) & $BR(B \to \tau \nu)_{ALPHA}= 1.065(21) \times 10^{-4}$  \\ 
    		&			&	  & $BR(B \to \tau \nu)_{exp}= 1.05(25) \times 10^{-4}$  \\ \hline
    $f_{B_s}$ [MeV]& 224(13) 	& 234(6) & $BR(B_s \to \mu^+ \mu^-)_{ALPHA}$ $=3.15(27) \times  10^{-9}$\hspace{-0.2cm}\\ 
    		   &		&	  & $BR(B_s \to \mu^+ \mu^-)_{exp}=2.9(0.7) \times 10^{-9}$ \\ \hline
    $f_{B_s}/f_B$ & $1.195 (61)(20)_{ChPT}\hspace{-0.2cm}$ & 1.19(05) &  \\ \hline
    \end{tabular}
    \caption{\label{tabpre} Comparison of ALPHA results for decay constants with FLAG 
    (see  \emph{http://itpwiki.unibe.ch/flag/index.php}) 
    results and experiment. In the last column we plug our values for $f_B$ and $f_{B_s}$ 
    into the SM predictions for $BR(B \to \tau \nu)$ and $BR(B_s \to \mu^+ \mu^-)$ respectively, 
    and compare with experiment. In the first 
    case we need to also input $|V_{ub}|$ from 
    the PDG 12, which is an average of the determination from 
    $BR(B \to \pi l \nu)$ and the one from inclusive decays; 
    in the latter we need to input $|V_{tb}^*V_{ts}|$ from 
    the CKM-fit (see  \emph{http://ckmfitter.in2p3.fr/}) , which therefore depends on a large set of 
    observables but mainly the $B_s^0$ splitting.}
\end{center}
\end{table}

Our results are summarized in Tables \ref{tabpost} 
and \ref{tabpre} together with a comparison with results from
other collaborations and experiment, while the quality of our extrapolations is illustrated in
Fig.~\ref{fBextrap} for some of our observables. 
More details about our computations, as well as 
a separation of $1/m_b$ effects, will appear in future publications.

\section{Conclusions}

As stressed in Table~\ref{tabpost} our method, based on NLO lattice HQET for the b-quark,
 requires 
just one input from experiment (chosen to be $m_B$) and
allows to make predictions for a wide class of observables. 
Our computation is the first unquenched one (although
with $N_f=2$) in which renormalization and matching were 
perfomed entirely in a non-perturbative way. Special
attention has been paid to the treatment of correlations 
and autocorrelations in the statistical analysis, and to
the treatment of excited states in the extraction of 
large-volume matrix elements. The agreement found for the
B-meson mass spectrum, albeit with larger errors, 
gives additional confidence in the correctness of the method
and in the values obtained for the decay constants. 
In the future these methods will be applied to 2+1-flavor simulations.

Concerning the b-quark mass it is quite surprising 
that our value agrees with the PDG one, considering the different systematics involved 
($N_f=2$ in our case). 
The error budget analysis shows that nearly $60\%$ of the statistical 
error comes from the determination of the HQET parameters 
$\mathbf{\omega}(z)$, indicating thus a clear path to 
further increase the precision of our results.

\vspace{0.3cm}

{\bf Acknowledgements}
{
\footnotesize This work is supported in part by the grants SFB/TR9, SFB~1044~(G.v.H.), HE~4517/2-1(P.~F. and J.~H.), 
and HE~4517/3-1 (J.~H.) of the Deutsche Forschungsgemeinschaft. We
are grateful for computer time allocated for our project on the Jugene 
and Juropa computers at NIC, J\"ulich, and the ICE at ZiB, Berlin. 
This work was granted access to the HPC resources of the Gauss
Center for Supercomputing at Forschungzentrum J\"ulich, Germany, made available
within the Distributed European Computing Initiative by the PRACE-2IP, receiving 
funding from the European Community's Seventh Framework Programme
(FP7/2007-2013) under grant agreement RI-283493.
}

\bibliography{proc13}{}
\bibliographystyle{unsrt}

\end{document}